\documentclass [prc,aps,showpacs]{revtex4}
\usepackage{graphicx}
\begin{document}

\newcommand{\pd}[2]{\frac{\partial #1}{\partial #2}}
\newcommand{\pdline}[2]{{\partial #1}/{\partial #2}}
\newcommand{\be}{\begin{equation}}
\newcommand{\ee}[1]{ \label{#1}  \end{equation}}

\newcommand{\adag}{a^{\dag}}
\newcommand{\atil}{\tilde{a}}
\def\frp*1{${*1\over2}^+$}
\def\frm*1{${*1\over2}^-$}
\def\g{\noindent}
\def\mev{\hbox{\MeV}}
\def\kev{\hbox{\keV}}
\def\lambdabar{{\mathchar'26\mkern-9mu\lambda}}
\def\lambdabarrr{{^-\mkern-12mu\lambda} }

\title{Thermodynamical limit in non-extensive and R\'{e}nyi statistics}
\author{A.S.~Parvan$^{a,b}$  and  T.S.~Bir\'{o}$^{c,}$}

\affiliation{$^{a}$Joint Institute for Nuclear Research, BLTP,
141980 Dubna, Russia}

\affiliation{$^{b}$Institute of Applied Physics, Moldova Academy
of Sciences, MD-2028 Chisinau, Republic of Moldova}

\affiliation{$^{c}$
KFKI Research Institute for Particle and
Nuclear Physics, H-1525 Budapest, P.O.Box 49, Hungary
}



\begin{abstract}
Previous results on R\'{e}nyi and Wang's formalism of the Tsallis
thermostatics are founded by using an
extensive variable $z$ connected to the entropic parameter $q$.
It is shown that in the thermodynamical limit both
thermostatics
meet all the requirements of equilibrium thermodynamics.
In particular, both the Tsallis and R\'{e}nyi entropies are
extensive functions of state and the temperature of the system is
intensive. In the thermodynamical limit Wang's incomplete
nonextensive statistics resembles the Tsallis one, but
the R\'{e}nyi thermostatics is reduced to the usual
Boltzmann-Gibbs one. The principle of additivity and the zeroth
law of thermodynamics in the canonical ensemble for both
thermostatics are demonstrated on the particular example of the
classical ideal gas of identical particles.
\end{abstract}

\maketitle

\section{Introduction}
The statistical mechanics based on the nonextensive Tsallis
entropy~\cite{Tsal88,Tsal98} finds support in recent studies
due to non-Gibbs phase distribution functions which appear
to be successful in analyzing certain experimental data~\cite{Tsal99,Gudima}.
In these investigations the statistical Boltzmann-Gibbs entropy is modified
by the additional parameter $q$, appearing both in the
distribution function and in the equation of state of the system.
They enlarge the range of application of the usual statistical mechanics
without taking into account realistic interactions in the microscopic
Hamiltonan.  In this respect two problems occur: the problem of the physical
interpretation of the parameter $q$ and the problem of connecting
such statistical treatments to equilibrium thermodynamics~\cite{Vives}.
The latter gives rise to
theoretical discussions in the literature due to the difficulties
in the proof of the zeroth law of thermodynamics. It is closely
connected to the principle of additivity. In the Tsallis
statistics the principle of additivity is violated as the
statistical entropy is nonextensive due to its definition. Therefore, in
the case when the parameter $q$ is a universal constant all
attempts to proof the zeroth law of thermodynamics for finite
systems failed~\cite{Abe0,Wang1,Abe1,Wang2,Abe2004}.
Moreover, in this case the thermodynamical limit is incompatible
with the Gibbs limit, $q\to 1$~\cite{Abe2}. It is well known
that equilibrium thermodynamics is a macroscopic,
phenomenological theory defined only in the thermodynamical
limit~\cite{Kvasn}. Therefore, an unambiguous connection between
statistical mechanics and equilibrium thermodynamics can
be provided only in the thermodynamical limit, while boundary
effects can be neglected. Thus, first of all, the correct
definition of the thermodynamical limit for the Tsallis statistics
and the interpretation of the parameter $q$ must be clarified.

A correct definition of the thermodynamical limit for the Tsallis
thermostatics was given first in Botet et al.~\cite{Botet1,Botet2}.
Later, in~\cite{Parv1}, it was revealed that if the parameter $1/(q-1)$ can
be interpreted as an extensive variable of state,
then in the microcanonical ensemble the Tsallis entropy  becomes
extensive in the thermodynamical limit. Thus, the zeroth law of
thermodynamics and the principle of additivity are restored,
all functions of state are either extensive or
intensive in conformity with the requirements of  equilibrium
thermodynamics. For the canonical ensemble similar results were
obtained in the framework of the classical ideal gas of
identical particles only~\cite{Parv2}, because in this case the
functions of state can be analytically integrated, and so exact
results can be obtained. Note that the procedure stated
in~\cite{Parv1,Parv2} assumes the regularization of the final
integrated expressions of the functions of state, while they depend
only on the variables of state of the system by applying the
thermodynamical limit condition. It allows to establish an
unambiguous connection of statistical mechanics to
equilibrium thermodynamics and to explore the validity of this
statistical mechanics using a given entropy.

In our previous paper~\cite {ParvBiro} the  microcanonical and
canonical ensembles of the Wang's formalism for the Tsallis
and the corresponding ensembles for the R\'{e}nyi entropy
were compared. It was shown that in Wang's formalism for the
Tsallis statistics both in the microcanonical and the canonical ensembles
the zero-th law of thermodynamics and the principle of additivity are
violated, as long as the parameter $q$ were a universal constant.
This conclusion is also valid for the R\'{e}nyi statistics in the canonical
ensemble. The microcanonical ensemble of the R\'{e}nyi statistics
on the other hand coincides with the Gibbs one.
Hence, all laws of the equilibrium thermodynamics are satisfied.
In the present paper we show that if the entropic parameter $q$
is connected with an extensive variable of state $z$
in a certain way, then in the thermodynamic
limit~\cite{Parv1,Parv2} both the R\'{e}nyi and  Wang's
incomplete nonextensive statistics become extensive and they
satisfy all laws of equilibrium thermodynamics.

This paper is organized as follows. In the second and third
sections our results for the microcanonical and canonical ensembles
for Wang's incomplete nonextensive statistics and R\'{e}nyi one are given.
In the thermodynamical limit the connection from statistical mechanics
to thermodynamics is implemented.

\section{Incomplete nonextensive statistics}

The incomplete nonextensive thermostatics or the Wang's
formalism of the generalized statistical mechanics is based on
Tsallis' definition of the statistical entropy~\cite{Tsal88}, and on the
incomplete normalization condition for the phase distribution
function~\cite{Wang00,Wang1},
\begin{equation}\label{1}
S = -k \int d\Gamma \frac{\varrho-\varrho^{q}}{1-q},
\;\;\;\;\;\;\;\; \int d\Gamma \varrho^{q}=1.
\end{equation}
This is conform with a modified expectation value of a dynamical variable $A$
\begin{equation}\label{2}
 \langle A\rangle = \int d\Gamma A \varrho^{q}.
\end{equation}
Here $\varrho^{q}$ is the phase distribution function, $d\Gamma
=dxdp$ is an infinitesimal element of phase space, $k$ is the
Boltzmann constant and $q\in\mathbf{R}$ is a real parameter,
$q\in [0,\infty]$. The main results in this formalism for the
microcanonical and canonical ensembles were obtained by us in
detail in~\cite{ParvBiro}. It was shown that if the parameter $q$
is a universal constant, i.e. it takes identical values for
different subsystems, then the Tsallis thermostatics violates
the zeroth law of thermodynamics and the connection between
statistical mechanics and equilibrium thermodynamics is lost.
Now, following the method defined in~\cite{Parv1,Parv2},
we  show that this violation can be removed. Note that
we  use  general relations obtained in our previous work~\cite{ParvBiro}.

\subsection{Microcanonical ensemble $(E,V,z,N)$}

Let us consider an isolated thermodynamical system, specified by $(E,V,z,N)$.
It is supposed that all dynamical systems in an equilibrium
statistical ensemble have identical energy $E$ within $\Delta E\ll
E$. The phase distribution function and the statistical weight in
the classical statistical mechanics can be written
as~\cite{ParvBiro}
\begin{eqnarray} \label{3}
 f &=& \varrho^{\frac{z}{z+1}}= W^{-1}\Delta(H-E), \\ \label{4}
 W &=& \int\Delta(H-E) d\Gamma,
\end{eqnarray}
where $H$ is the Hamilton function and $\Delta(\varepsilon)$ is
the function distinct from zero only in the interval
$0\leq\varepsilon\leq\Delta E$, where it is equal to unity. The
thermodynamical variable of state $z$ is expressed through the
parameter $q$~\cite{Parv1} as
\begin{equation}\label{5}
 z=\frac{q}{1-q}.
\end{equation}
Then, the entropy of the system in terms of the variables of
state $(E,V,z,N)$ takes the following form~\cite{ParvBiro}
\begin{equation}\label{6}
     S = k(z+1) [1-e^{-S_{G}/kz}],
\end{equation}
where $S_{G}=k\ln W$ is the usual Boltzmann-Gibbs entropy in the
microcanonical ensemble.


In the microcanonical ensemble the division of a total system into
two dynamically independent subsystems, $H=H_{1}+H_{2}$, for
finite number of particles, $N=N_{1}+N_{2}$, leads to a
convolution of the statistical weight $W$
\begin{equation}\label{1c}
    W(E,V,N)=\frac{N_{1}!N_{2}!}{N!} \int\limits_{0}^{E} dE_{1}
    W_{1}(E_{1},V,N_{1}) W_{2}(E-E_{1},V,N_{2}),
\end{equation}
where $W_{i} = \int\Delta(H_{i}-E_{i}) d\Gamma_{i}$. The
statistical weight (\ref{4}) does not factorize for finite energy
$E$ and finite number of particles $N$ in a finite volume $V$.
Hence, in the microcanonical ensemble even the Gibbs entropy is
nonadditive, $S_{G}\neq S_{G,1}+S_{G,2}$, and consequently the
Tsallis entropy is also nonadditive, $S \neq S_{1}+S_{2}$. Only in
the thermodynamic limit ($N\to \infty,E\to\infty,V\to\infty$ and
$\varepsilon=E/N=const,v=V/N=const$) the statistical weight
(\ref{1c}) factorizes, $W=W_{1}W_{2}$. Factorizing requires the
$N$-dependence $W=w^N$ with $w=w(\varepsilon,v)$ a function
depending only on intensive variables of state. Then, the Gibbs
entropy is additive, but for the Tsallis entropy there exist two
possibilities: the parameter $z$ can be either intensive or
extensive.


First we present arguments in favor of treating the parameter $z$ as
an extensive variable. We inspect the equality of temperatures of
two subsystems kept in equilibrium in the thermodynamical limit both
by assuming a universal and an extensive value $z=N\tilde{z}$.

The inverse temperature, by definition, is given as
$1/T=\pdline{S}{E}$. For the Tsallis definition of entropy it
becomes \be\label{2c}
 \frac{1}{T} \: = \: k \frac{z+1}{z} W^{-1/z} \, \pd{\ln W}{E}.
\ee{INV_TEMPERATURE} In the thermodynamical limit $W=W_1W_2$ and
$N=N_1+N_2$ while specific values, like $\varepsilon=E/N$, are
constant. The inverse temperature in this limit becomes
\be\label{3c}
 \frac{1}{T} \: = \: k \frac{z+1}{z} w^{-N/z} \, \pd{\ln w}{\varepsilon}.
\ee{LIMIT_INV_TEMP} Considering $z$ as a universal constant, we
arrive at an $N$-dependent temperature even in the thermodynamical
limit, and therefore the equality of temperatures between large
subsystems is spoiled: $T_1 \ne T_2 \ne T$. Moreover, the Tsallis
entropy is nonadditive
\begin{equation}\label{4c}
    S=S_{1}+S_{2}-\frac{1}{k(z+1)} S_{1}S_{2}.
\end{equation}
On the contrary if $z$ is treated as an extensive variable itself,
$z = N \tilde{z}$, then in the $N \rightarrow \infty$ limit
equilibrium is established via
\begin{equation}\label{5c}
 T_1 = T_2 = T, \qquad \tilde{z}_1 = \tilde{z}_2 = \tilde{z}
\end{equation}
and $S = S_{1}+S_{2}$. One concludes that treating $z$ as
extensive the Tsallis entropy is additive in the thermodynamic
limit and the zero law of thermodynamics is satisfied.


Investigating the thermodynamical functions of state in the
thermodynamical limit amounts to expanding these functions in
inverse powers of a large extensive parameter, for example, $1/N$
($N\gg 1, |z|\gg 1, E\gg 1,V\gg 1$ and
$\varepsilon=E/N=const,v=V/N=const,\tilde{z}=z/N=const$), while
keeping ratios to $N$ (specific values) finite~\cite{Parv1}. In the
thermodynamic limit the entropy~(\ref{6}) for large values of
$|z|\gg 1$ can be written as
\begin{equation}\label{7}
 S = k z [1-e^{-S_{G}/kz}].
\end{equation}
The Boltzmann-Gibbs entropy is an extensive function, $S_{G}
(\lambda E, \lambda V, \lambda N) = \lambda S_{G} (E, V, N)$.
Then, the Tsallis entropy in the microcanonical ensemble~(\ref{7})
is a first order homogeneous function of the variables $E,V,z,N$
\begin{equation} \label{8}
 S (\lambda E, \lambda V, \lambda z,
\lambda N) = \lambda S (E, V, z, N).
\end{equation}
A complete account of the derivation of
thermodynamical relations based on Eq.~(\ref{8}) can be found
in~\cite{Parv1}. Here we augment only the main results. First,
Euler's theorem for homogeneous functions can be applied to
Eq.~(\ref{8}). We obtain
\begin{equation} \label{9}
 T S=E+p V +X z-\mu N.
\end{equation}
Then the partial derivatives are given as
\begin{eqnarray} \label{10}
    \left(\frac{\partial S} {\partial E} \right)_{V, z, N} &=&
     \frac{1} {T}, \;\;\;\;\;\;
     \left(\frac{\partial S} {\partial
     V} \right)_{E, z, N} = \frac{p} {T}, \nonumber \\
      \left(\frac{\partial S} {\partial z} \right)_{E, V, N} &=& \frac{X} {T},
 \;\;\;\;\;\;
     \left(\frac{\partial S} {\partial
     N} \right)_{E, V, z} =-\frac{\mu} {T}.
\end{eqnarray}
and finally Euler's theorem provides the fundamental differential form,
\begin{equation} \label{11}
  T dS = dE + p dV +X dz - \mu dN,
\end{equation}
as well as the Gibbs-Duhem relation
\begin{equation}\label{12}
    S dT = V dp +z dX-N d\mu,
\end{equation}
and consequently the first and the second laws of macroscopic
thermodynamics. The temperature $T$, the variable $X$, the
pressure and the chemical potential of the system are expressed
through the variables of the Gibbs statistics, as follows:
\begin{eqnarray} \label{13}
  T &=& T_{G} \ e ^{S_{G}/k z}, \\ \label{14}
  X &=& k T_{G}[e ^{S_{G}/k z} - (1+S_{G}/k z)]
\end{eqnarray}
and $p = p_{G}$, $\mu=\mu_{G}$, respectively. Thus, the
temperature $T$, the pressure $p$, the chemical potential $\mu$,
and the quantity $X$ are homogeneous functions of the variables
$E,V,z,N$  of order zero: They are intensive variables. Note that
in the thermodynamical limit all relations in the microcanonical
ensemble concerning Wang's formalism of Tsallis thermostatics
completely coincide with the ones of the original Tsallis
thermostatics given in~\cite{Parv1}. Moreover, all conclusions
relative to the Tsallis statistics in the microcanonical ensemble
given in~\cite{Parv1} are also preserved in the case of the Wang's
formalism in the thermodynamical limit. Therefore, whenever the
parameter $z=q/(1-q)$ is an extensive variable of state, then in
the thermodynamical limit the principle of additivity and the
zeroth law of thermodynamics are valid. In particular the Tsallis
entropy is extensive and the temperature is intensive. In this
case the Tsallis statistics completely satisfies all postulates of
equilibrium thermodynamics. An evaluation of thermodynamical
identities in the microcanonical ensemble is provided by the Euler
theorem.

In the thermodynamical limit in the microcanonical ensemble
results for an ideal gas of identical particles are coincident in
both Wang's and Tsallis' original formalisms. Expressions for the
ideal gas in the microcanonical ensemble for the Tsallis
statistics can be found in~\cite{Parv1}.
\begin{figure}
  \includegraphics[width=16cm]{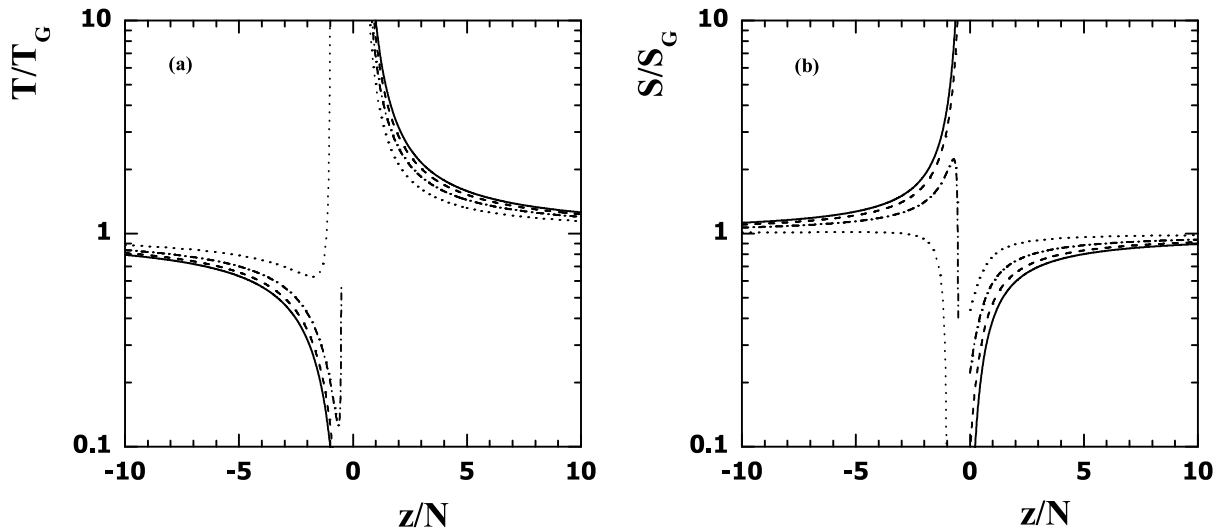}
  \caption{The dependence of the ratio of temperature $T$ to the Gibbs temperature $T_{G}$
  (a) and the ratio of the entropy $S$ to the Gibbs entropy $S_{G}$ (b) on
  the specific $\tilde{z}=z/N$ for the classical non-relativistic ideal gas of nucleons in
  the microcanonical ensemble at the
values of the specific energy $\varepsilon=50$ MeV and the
specific volume $v=3/\rho_{0}$. The solid lines represent the
calculations treating $z$ as extensive. The dotted, dash-dotted
and dashed lines correspond to the calculations considering $z$ as
a universal constant for the number of particles $N=1,2$ and $5$,
respectively. }\label{fg1}
\end{figure}
Here we present relations for the temperature and entropy either
considering the parameter $z$ as a universal constant or as a
extensive variable of state. The one-particle statistical weight
for the classical non-relativistic ideal gas is given by
\begin{equation}\label{6c}
    w=v\left( \frac{m\varepsilon
    e^{5/3}}{3\pi\hbar^{2}}\right)^{3/2}.
\end{equation}
For a universal $z=z_{1}=z_{2}$ in the thermodynamical limit the
temperature (\ref{3c}) and entropy (\ref{6}) are given by
\begin{eqnarray}
  T &=& \frac{2}{3} \frac{\varepsilon}{k}\ \frac{z}{z+1} \ w^{N/z}, \\
  S &=& k(z+1)\left[ 1-w^{-N/z}\right].
\end{eqnarray}
This example illustrates that in the original Tsallis' approach
the temperature is not-intensive and entropy is not-extensive. On
the other hand if the parameter $z$ is extensive, $z=N\tilde{z}$,
then in the thermodynamical limit ($N\gg 1$) the temperature (\ref{3c}) and
entropy (\ref{6}) take the form
\begin{eqnarray}
  T &=& \frac{2}{3} \frac{\varepsilon}{k} \ w^{1/\tilde{z}}, \\
  S &=& k \tilde{z}N \left[ 1-w^{-1/\tilde{z}}\right].
\end{eqnarray}
In this case the temperature is intensive and the entropy is
extensive, in complete agreement with requirements of equilibrium
thermodynamics.


\subsection{Canonical ensemble $(T,V,z,N)$}
In the canonical ensemble in Wang's formalism of Tsallis
thermostatics the equilibrium phase distribution function can
be written as~\cite{ParvBiro}
\begin{equation}\label{15}
\varrho^{\frac{z}{z+1}}=\left[1+\frac{z}{(z+1)^{2}}\frac{\Lambda-H}{kT}\right]^{z},
\end{equation}
where $\Lambda=\langle H\rangle-\frac{z+1}{z}\ TS$ is determined
from the normalization condition (\ref{1})
\begin{equation}\label{16}
\int \left[1+\frac{z}{(z+1)^{2}}\frac{\Lambda-H}{kT}\right]^{z}
d\Gamma=1.
\end{equation}
Thus $\Lambda$, by solving Eq.~(\ref{16}), is a function of the
variables of state, $\Lambda =\Lambda(T,V,z,N)$. The expectation
value $\langle A \rangle$ of a dynamical variable $A$ can be
computed as (cf.~(\ref{2})):
\begin{equation}\label{17}
\langle A\rangle =\int A
\left[1+\frac{z}{(z+1)^{2}}\frac{\Lambda-H}{kT}\right]^{z}
d\Gamma.
\end{equation}
The entropy takes the form
\begin{equation}\label{18}
      S=\frac{z}{z+1}\ \frac{\langle H\rangle -\Lambda}{T},
\end{equation}
and the free energy is unambiguously determined by the functions
$\Lambda$ and $\langle H\rangle$ as the Legendre transform of
energy with respect to the entropy of the system
\begin{equation}\label{19}
      F\equiv \langle H\rangle -TS=\frac{\langle H\rangle+z
    \Lambda}{z+1} .
\end{equation}


In the canonical ensemble of Gibbs statistics the division of a
total system into two dynamically independent subsystems,
$H=H_{1}+H_{2}$, for finite number of particles, $N=N_{1}+N_{2}$,
does not lead to factorization of the distribution function,
$\varrho_{G}\neq \varrho_{G,1}\varrho_{G,2}$, due to the relation
\begin{equation}\label{1d}
    Z(T,V,N)=\frac{N_{1}!N_{2}!}{N!}
    Z_{1}(T,V,N_{1})Z_{2}(T,V,N_{2}),
\end{equation}
where $Z=\int d\Gamma \exp(-H/T)$ and
$\varrho_{G}=Z^{-1}\exp(-H/T)$ are the canonical partition
function and the distribution function, respectively. The
partition function (\ref{1d}) does not factorize because of the
Gibbs factor $N!$ and finite volume $V$. In the canonical ensemble
even the Gibbs entropy is not-additive for finite systems, $S_{G}\neq S_{G,1}+S_{G,2}$.
Hence for dynamically independent subsystems at finite values of
$N,V$ we have statistical dependence. In case of Tsallis
statistics we have equivalent results for the distribution
function, $\varrho\neq \varrho_{1}\varrho_{2}$, and entropy $S\neq
S_{1}+S_{2}$ at finite values of $N,V$. In the thermodynamical
limit ($N\to\infty,V\to\infty,v=const$) the Gibbs and Tsallis
approaches differ essentially. The Gibbs partition function
factorizes, $Z_{G}=Z_{G,1}Z_{G,2}$, it leads to
$Z_{G}=\tilde{Z_{G}}^{N}$, where
$\tilde{Z_{G}}=\tilde{Z_{G}}(T,v)$ is an intensive function of
intensive variables of state only. For the Tsallis statistics
there are two possibilities. First, if the parameter $z$ is a
universal constant then even in the thermodynamic limit
($N\to\infty,V\to\infty,v=const,z=const$) the distribution
function (\ref{15}) does not factorize
\begin{equation}\label{2d}
\varrho^{\frac{z}{z+1}}\neq
\varrho_{1}^{\frac{z}{z+1}}\varrho_{2}^{\frac{z}{z+1}}.
\end{equation}
The entropy is not-additive and Tsallis formula (\ref{4c}) cannot
be applied. On the other hand, keeping $\tilde{z}=z/N$
constant in the thermodynamic limit, two dynamically independent
subsystems are independent also statistically. In this case the
distribution function (\ref{15}) can be written as
\begin{equation}\label{3d}
    \varrho=\left[1+\frac{1}{\tilde{z}}
    \frac{\lambda - \tilde{H}}{T}\right]^{\tilde{z}N},
\end{equation}
where $\lambda=\Lambda/N$ is an intensive variable,
$\lambda_{1}=\lambda_{2}=\lambda$. If the reduced Hamiltonian
(energy per particle),
$\tilde{H}=H/N$, is assumed to be an intensive function,
$\tilde{H}_{1}\sim\tilde{H}_{2}\sim\tilde{H}$, then for equal
temperatures, $T_{1}=T_{2}=T$, the distribution function
(\ref{3d}) in fact factorizes and the entropy is additive
\begin{eqnarray}\label{4d}
\varrho &=&  \varrho_{1}\varrho_{2}, \;\;\;\;\;\;\;\;\;\;\;
\tilde{z}=\tilde{z}_{1}=\tilde{z}_{2}, \\
S&=& S_{1}+S_{2}.
\end{eqnarray}
Note that the distribution function (\ref{3d}) is expressed in the
form, $\varrho=\tilde{\varrho}^{N}$.

\begin{figure}
  \includegraphics[width=17cm]{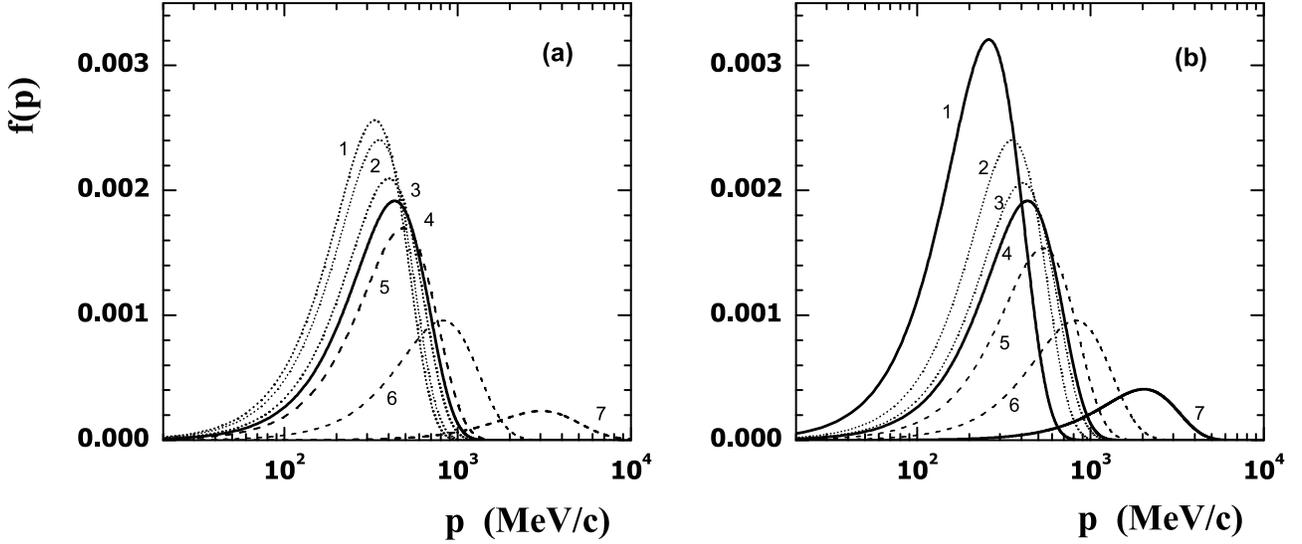}
  \caption{The single-particle distribution function for classical non-relativistic
  ideal gas of nucleons in Wang's formalism of Tsallis statistics at the
temperature $T=100$ MeV and the specific volume $v=3/\rho_{0}$ for
two treatments of a parameter $z$: (a) a universal
constant~\cite{ParvBiro} for $N=10$ and different values of
$z=20$, $30$, $100$, $\pm\infty$, $-100$, $-30$ and $-20$ (the
curves $1,2,3,4,5,6$ and $7$, respectively), (b) an extensive
variable of state~\cite{Parv2} in thermodynamic limit for
different values of the specific $\tilde{z}=3$, $\pm\infty$, $-3$
(solid lines $1,4,7$, respectively). The dotted and dashed lines
(b) correspond to the calculations considering $z$ as an universal
constant~\cite{ParvBiro} for $\tilde{z}=3$ and number of particles
$N=10,50$ (lines 2,3) and $\tilde{z}=-3$ and $N=50,10$ (lines
5,6). The line 4 on both panels corresponds to the conventional
Boltzmann-Gibbs statistics. }\label{fg2}
\end{figure}

Let us now deduce the fundamental equation of thermodynamics for
Tsallis statistics with an extensive $z$. Applying the total
differential operator with respect to all ensemble variables
$(T,V,z,N)$ on the entropy $S$, the energy $\langle H \rangle$ and
the norm equation (\ref{1}), and using Eq.~(\ref{15}) and the
parametric dependence of the Hamilton function $H$ on the
variables $V$, $N$ and $z$ one finds the fundamental equation of
thermodynamics~\cite{Parv2,Vives,ParvBiro}
\begin{equation}\label{23}
     T dS= d\langle H\rangle + pdV +X dz -\mu dN,
\end{equation}
where
\begin{eqnarray}\label{24}
  p &=& \int d\Gamma \varrho^{\frac{z}{z+1}}
 \left(-\frac{\partial H}{\partial V}
 \right)_{T,z,N}  =\left\langle -\frac{\partial H}{\partial V}
 \right\rangle, \\ \label{25}
  \mu &=& \int  d\Gamma \varrho^{\frac{z}{z+1}}
 \left(\frac{\partial H}{\partial N}
 \right)_{T,V,z} =\left\langle \frac{\partial H}{\partial N}
 \right\rangle, \\ \label{21}
 X &=& \int d\Gamma \varrho^{\frac{z}{z+1}}
 \left\{kT\left[1-\varrho^{\frac{1}{z+1}}
 \left(1-\frac{z+1}{z}\ln\varrho^{\frac{1}{z+1}}\right)\right]
 + \left(-\frac{\partial H}{\partial z}\right)_{T,V,N}\right\}.
\end{eqnarray}
Here, the property of the Hamilton function $(\partial H/\partial
T)_{V,z,N}=0$ was assumed. Using Eqs.~(\ref{19}) and (\ref{23}) we
get a differential formula for the free energy
\begin{equation}\label{26}
    dF= -S dT - p dV- X dz + \mu dN.
\end{equation}

In the canonical ensemble the thermodynamical limit means to let
$N\rightarrow\infty$, while $v=V/N$ and $\tilde{z}=z/N$ kept constant.
We make an expansion of functions of state in powers of the small parameter $1/N$
(with $N\gg 1$ it is $|z|\gg 1$ and $V\gg 1$)~\cite{Parv2}.
Note that in the
thermodynamical limit all relations of the canonical ensemble
derived here completely coincide with ones of the original Tsallis
thermostatics given in~\cite{Parv2}. Moreover, all results for
the ideal gas in the canonical ensemble for both formalisms are
equivalent. To supply evidence, we should rewrite the expressions for the perfect
gas in the canonical ensemble in Wang's formalism given in~\cite{ParvBiro}
in terms of the variable $z$ (\ref{5}),
and then impose the thermodynamical limit.
Expressions for the ideal gas in the canonical ensemble for the Tsallis statistics
can be found in~\cite{Parv2}. Thus all conclusions relative to
Tsallis statistics in the canonical ensemble given
in~\cite{Parv2} are preserved in the case of Wang's
formalism in the thermodynamical limit.
The statistical mechanics based on Tsallis entropy completely satisfies all requirements
of equilibrium thermodynamics in the canonical ensemble in the thermodynamical limit.
It was shown that all functions of state of the system are the homogeneous
functions of the first degree, extensive, or the homogeneous
functions of the zero degree, intensive. In particular, the
temperature is an intensive variable and thus provides
implementation of the zero law of thermodynamics.

\section{R\'{e}nyi thermostatics}

The R\'{e}nyi thermostatics is based on R\'{e}nyi's
definition of statistical entropy with a usual norm equation for
the phase distribution function
\begin{equation}\label{30}
    S=k\frac{\ln\left(\int \varrho^{q} d\Gamma\right)}{1-q},
    \;\;\;\;\;\;\;\;\;\;\;\; \int \varrho d\Gamma = 1.
\end{equation}
The expectation value of a dynamical variable $A$ is given as
\begin{equation}\label{31}
 \langle A\rangle = \int A \varrho d\Gamma.
\end{equation}
Note that the main relations for R\'{e}nyi thermostatics
were derived in~\cite{ParvBiro}. Here our investigation proceeds
with the description of the entropic parameter $q$ through a
thermodynamical variable of state $z$,
\begin{equation}\label{32}
  z=\frac{1}{q-1}.
\end{equation}
This allows us to introduce a correct thermodynamical limit
consistent with the macroscopic laws of equilibrium
thermodynamics.
We rewrite all expressions of the R\'{e}nyi thermostatics given in~\cite{ParvBiro}
in terms of the variable $z$ and implement the thermodynamical limit
described above.

\subsection{Microcanonical ensemble $(E,V,z,N)$}

In the microcanonical ensemble of the R\'{e}nyi thermostatics
the phase distribution function and the statistical weight
are~\cite{ParvBiro}
\begin{eqnarray} \label{33}
 \varrho &=& W^{-1}\Delta(H-E), \\ \label{34}
 W &=& \int\Delta(H-E) d\Gamma.
\end{eqnarray}
Then, the R\'{e}nyi entropy (\ref{30}) is reduced to the familiar
expression
\begin{equation}\label{35}
    S=k\ln W \equiv S_{G},
\end{equation}
where $S_{G}$ is the Gibbs entropy which does not depend on the
variable of the state $z$. Hence, all expressions of the R\'{e}nyi
statistics in the microcanonical ensemble are equivalent with ones
of the conventional Gibbs statistics. The conclusions
concerning the microcanonical ensemble of the corresponding Gibbs
thermostatics can be generalized to apply to the R\'{e}nyi one.
The R\'{e}nyi entropy in the microcanonical
ensemble~(\ref{35}) is a first order homogeneous function of variables $E,V,N$
\begin{equation} \label{35a}
 S (\lambda E, \lambda V,
\lambda N) = \lambda S (E, V, N).
\end{equation}
Eq.~(\ref{35a}) provides the Euler theorem, the fundamental
equation of thermodynamics, the Gibbs-Duhem relation and
consequently the first and the second laws of the macroscopic
thermodynamics which all are independent of the parameter $q$. All
functions of state in the thermodynamical limit are either
intensive or extensive according to requirements of equilibrium
thermodynamics. Thus, the principle of additivity and the zero law
of thermodynamics are valid. The R\'{e}nyi statistics in the
microcanonical ensemble completely satisfies all requirements of
the equilibrium thermodynamics.  Note that specific expressions
for the ideal gas in the microcanonical ensemble for the Gibbs
statistics can be found in~\cite{Parv1}.

\subsection{Canonical ensemble $(T,V,z,N)$}
In the canonical ensemble of the R\'{e}nyi thermostatistics the
phase distribution function depends on two, directly unknown,
variables $S$ and $\langle H\rangle$~\cite{ParvBiro}
\begin{equation}\label{36}
\varrho=e^{-S/k} \left[1+\frac{1}{z+1}\frac{\langle
H\rangle-H}{kT}\right]^{z}.
\end{equation}
In order to normalize the phase distribution function (\ref{36})
two equations have to be applied~\cite{ParvBiro}: $\int d\Gamma
\varrho =1$ and $\langle H\rangle=\int d\Gamma H \varrho$. We
obtain the entropy and the energy of the system as functions of
the variables of state, $S=S(T,V,z,N)$ and $\langle
H\rangle=\langle H\rangle(T,V,z,N)$, respectively. The free energy
can be written as
\begin{equation}\label{39}
    F=\langle H\rangle-TS.
\end{equation}
The expectation value of a dynamical variable $A$ in this case is given by
\begin{equation}\label{40}
 \langle A\rangle=
e^{-S/k} \int A \left[1+\frac{1}{z+1}\frac{\langle
H\rangle-H}{kT}\right]^{z} d\Gamma.
\end{equation}

Let us derive the fundamental equation of thermodynamics at fixed
values of the variables of state $(T,V,z,N)$ for R\'{e}nyi
statistics with an extensive $z$. Following the procedure given
above by virtue of the parametric dependence of the Hamilton
function $H$ on the variables $V$, $z$ and $N$ only we obtain the
fundamental equation of thermodynamics~\cite{Vives,Parv2,ParvBiro}
\begin{equation}\label{43}
     T dS= d\langle H\rangle + pdV +X dz -\mu dN,
\end{equation}
with
\begin{eqnarray}\label{44}
  p &=& \int d\Gamma \varrho
 \left(-\frac{\partial H}{\partial V}
 \right)_{T,z,N}  =\left\langle -\frac{\partial H}{\partial V}
 \right\rangle , \\ \label{45}
  \mu &=& \int d\Gamma \varrho
 \left(\frac{\partial H}{\partial N}
 \right)_{T,V,z}  =\left\langle \frac{\partial H}{\partial N}
 \right\rangle, \\
 \label{42}
    X&=& kT\left[\frac{S}{kz}+e^{\frac{S}{kz}}
    \int d\Gamma \varrho^{\frac{z+1}{z}}\ln\varrho^{\frac{1}{z}}
    \right]+\int d\Gamma \varrho
 \left(-\frac{\partial H}{\partial z}
 \right)_{T,V,N}.
\end{eqnarray}
Here the property of the Hamilton function, $(\partial H/\partial
T)_{V,z,N}=0$, was used. For the R\'{e}nyi thermostatics
Eq.~(\ref{26}) is valid and the functions of state $X,p$ and $\mu$
can be calculated from the corresponding thermodynamical
relations. It is important to note that in the thermodynamical
limit all expressions of the ideal gas~\cite{ParvBiro} are exactly
equivalent with ones of the Gibbs statistics both the
thermodynamical functions of state and the one-particle
distribution function. In the last case in the thermodynamical
limit the non-relativistic single-particle distribution function
is reduced to the Maxwell-Boltzmann distribution $ f(\vec{p})
=(2\pi mkT)^{-3/2} \exp(-\vec{p}^{2}/2mkT)$ upon  rewriting the
expressions for the ideal gas in the canonical ensemble for the
R\'{e}nyi statistics in terms of the variable $z$ (\ref{32}), as
it is given in~\cite{ParvBiro}, and impose then the
thermodynamical limit condition. Specific expressions for the
ideal gas in the canonical ensemble for the Gibbs statistics can
be found in~\cite{Parv2}. Thus, all conclusions concerning the
canonical ensemble for the Gibbs statistics on the particular
example of the ideal gas can be generalized to apply to the
R\'{e}nyi thermostatics. The R\'{e}nyi statistics in the canonical
ensemble completely satisfies all requirements of equilibrium
thermodynamics. In particular, the principle of additivity, the
zeroth, the first, the second and the third laws of thermodynamics
are valid.


\section{Conclusion}

In this paper Wang's formalism of the Tsallis statistics was compared to
the R\'{e}nyi one in the thermodynamical limit.
An unambiguous connection between both statistical mechanicses and the
equilibrium thermodynamicses are revealed. For this purpose expressions
obtained in Wang's formalism of Tsallis statistics and the R\'{e}nyi one
were rewritten in terms of a new extensive variable of state, $z$,
related to the entropic parameter $q$. The functions of state were regularized
by applying the limiting procedure of the
thermodynamical limit. We obtained that in this limit
Wang's formalism of the Tsallis statistics in the terms of the
extensive variable of state $z$ completely coincide with the
original Tsallis thermostatics in both the microcanonical and
the canonical ensembles. However, the R\'{e}nyi statistics
resembles the usual Boltzmann-Gibbs thermodynamics.
In the microcanonical ensemble we proved for both the R\'{e}nyi
and Wang's statistics that in the thermodynamical limit
all laws of thermodynamics, in particular the zeroth law,
the principle of additivity, the Euler theorem, the
fundamental equation of thermodynamics and the Gibbs-Duhem
relation are valid. To put it simply both the Tsallis and the R\'{e}nyi
entropies are extensive in the thermodynamical limit, provided one
composes subsystems with the proper, and hence different, $q$-values.

In the canonical ensemble, however, only
the fundamental equation of thermodynamics, the first and the
second laws were derived in general terms.
For demonstrating further principles of
equilibrium thermodynamics in the framework of the canonical
ensemble,
both for the R\'{e}nyi statistics and for Wang's formalism of
Tsallis statistics,  exact analytical results were utilized for the ideal gas
of identical particles. It was shown for this
particular example that in the thermodynamical limit for both
thermostatics the main thermodynamical equations, the zeroth law
and the principle of additivity are satisfied. All functions of
state are either extensive or intensive. Moreover, in the canonical
ensemble in the thermodynamical limit both the Tsallis and the
R\'{e}nyi entropies are extensive.
For the ideal gas of identical particles in both thermostatics
the equivalence of the canonical and the microcanonical ensembles
in the thermodynamical limit was demonstrated. This is a very important property
to be verified for the self-consistent definition of any
statistical mechanics.

{\bf Acknowledgments:} This work has been supported by the
MTA-JINR Grant and OTKA T49466. One of authors, A.S.P.,
acknowledges valuable remarks and fruitful discussions with
R.~Botet, K.K.~Gudima, M.~P{\l}oszajczak and V.D.~Toneev.

\end{document}